\documentclass[sigconf,natbib=true,anonymous=false]{acmart}
\usepackage{multirow}
\usepackage{dsfont}
\newcommand{\PM}{MTLDS}
\newcommand{\DT}{eBay}

\AtBeginDocument{%
  \providecommand\BibTeX{{%
    \normalfont B\kern-0.5em{\scshape i\kern-0.25em b}\kern-0.8em\TeX}}}

\copyrightyear{2023}
\acmYear{2023}
\setcopyright{acmlicensed}\acmConference[SIGIR '23]{Proceedings of the 46th International ACM SIGIR Conference on Research and Development in Information Retrieval}{July 23--27, 2023}{Taipei, Taiwan}
\acmBooktitle{Proceedings of the 46th International ACM SIGIR Conference on Research and Development in Information Retrieval (SIGIR '23), July 23--27, 2023, Taipei, Taiwan}
\acmPrice{15.00}
\acmDOI{10.1145/3539618.3592023}
\acmISBN{978-1-4503-9408-6/23/07}

\begin{document}

\title{Modeling Orders of User Behaviors via Differentiable Sorting: A Multi-task Framework to Predicting User Post-click Conversion}


\author{Menghan Wang} 
\email{wangmengh@zju.edu.cn}
\affiliation{%
  \institution{eBay Inc.} 
  \city{Shanghai}
  \country{China} 
}
 
\author{Jinming Yang} 
\email{yangjm67@sjtu.edu.cn}
\affiliation{%
  \institution{Shanghai Jiaotong University} 
  \city{Shanghai}
  \country{China} 
}

\author{Yuchen Guo} 
\email{yuchguo@ebay.com}
\affiliation{%
  \institution{eBay Inc.} 
  \city{Shanghai}
  \country{China} 
}

\author{Yuming Shen} 
\email{yumishen@ebay.com}
\affiliation{%
  \institution{eBay Inc.} 
  \city{Shanghai}
  \country{China} 
}

\author{Mengying Zhu} 
\email{mengyingzhu@zju.edu.cn}
\affiliation{%
  \institution{Zhejiang University} 
  \city{Hangzhou}
  \country{China} 
}

\author{Yanlin Wang} 
\email{wangylin36@mail.sysu.edu.cn}
\affiliation{%
  \institution{Sun Yat-sen University} 
  \city{Zhuhai}
  \country{China} 
}

\renewcommand{\shortauthors}{Menghan Wang et al.}

\begin{abstract}
User post-click conversion prediction is of high interest to researchers and developers. Recent studies employ multi-task learning to tackle the selection bias and data sparsity problem, two severe challenges in post-click behavior prediction, by incorporating click data. However, prior works mainly foucsed on pointwise learning and the orders of labels (i.e., click and post-click) are not well explored, which naturally poses a listwise learning problem. Inspired by recent advances on differentiable sorting, in this paper, we propose a novel multi-task framework that leverages orders of user
behaviors to predict user post-click conversion in an end-to-end
approach. Specifically, we define an aggregation operator to combine predicted outputs of different tasks to a unified score, then we use the computed scores to model the label relations via differentiable sorting. Extensive experiments on public and industrial datasets show the superiority of our proposed model against competitive baselines.
 
\end{abstract}




\maketitle

\section{Introduction}
User post-click behaviors (e.g., purchase, download, and registration) reflect explicit preference of users and are often coherent with business metrics, attracting more research efforts nowadays to predict subsequent actions of users after they click some items. Early work revealed two non-trivial challenges when directly estimating post-click conversion (often abbreviated as CVR): 1) The sample selection bias \citep{zadrozny2004learning}. Conventional CVR models are trained on dataset composed of clicked impressions, while are utilized to make inference on the entire space with samples of all impressions. This inconsistency will hurt the generalization performance of trained models. 2) The data sparsity problem. In practice, post-click behaviors are much sparser than clicks, which would pose negative impacts on the effectiveness of CVR models. A common and natural solution is to embrace multi-task learning (MTL), i.e., learning click and post-click predictions at the same time. We could feed the whole dataset, without additional sampling for post-click prediction, into MTL models and share knowledge across tasks. Evidences \cite{ma2018entire,wen2020entire} showed MTL can help to circumvent the sample selection bias and alleviate the sparsity issue.


\begin{figure}[t]
\centering 
\includegraphics[width=8cm]{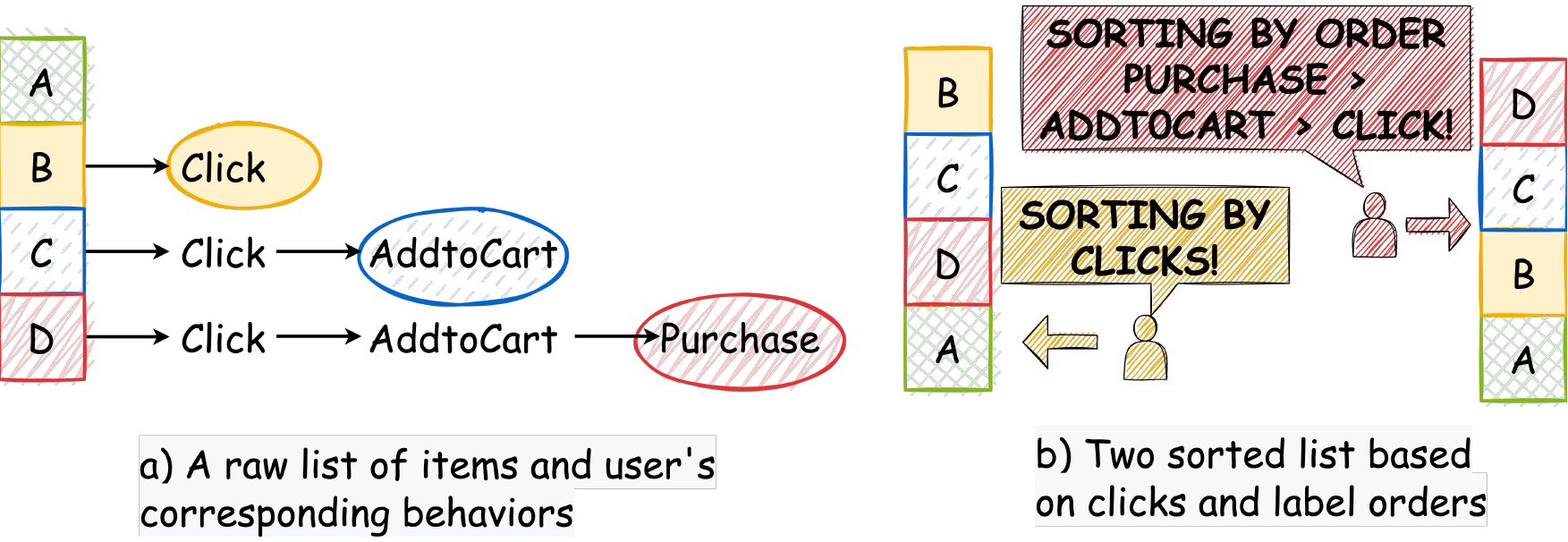}
\caption{An explainable example of motivation; a) a snapshot of user behaviors on a list of items, and b) two preference list sorted by pure clicks and label orders. }
\label{example}
\end{figure}

Then a subsequent issue arises under the MTL scenario is how to model the label relations of the multiple tasks. Ideally, a good answer should be 1) \textbf{scalable}. It is often the case that \textit{Post-click} is a subsequence since a user may take several actions after click, so the method should handle multiple labels as well as their predictions. For example in Fig. \ref{example}, \emph{Click -> AddtoCart -> Purchase} is a case with multiple post-click behaviors. To our best knowledge, there is no previous study that intentionally addressed this scalability issue; existing methods~\cite{ma2018entire,wen2020entire} that take probability multiplication may lead to a high-variance porblem when the number of labels increases.
2) \textbf{differentiable}. The merit of MTL relies on the end-to-end learning paradigm; by sharing the process of back propagation each task can benefit from other tasks. Similarly, the learning process of label relations should also be differentiable to fully leverage the potential of MTL. Moreover, we make a reasonable assumption: the order of user behavior follows the ordering of the label sequence. That is, in the above example user have to \textit{click} before \textit{AddtoCart}, and \textit{AddtoCart} before \textit{Purchase}; of course users can stop at any node in real scenarios. In this regard, the longer a user interacts an item along the sequence, the deeper engagement we can infer he/she builds with the item. Then, we can depict a relative preference order, i.e., \textit{Purchase} > \textit{AddtoCart} > \textit{Click}, which is more reasonable than the counterpart sorted by a single label (i.e., click) in Fig. \ref{example}. Thus we argue that sorting user preference by label orders may be a plausible answer to the label relation concern of MTL.



Building on the aforementioned insights, in this paper, we propose a scalable multi-task framework that leverages orders of user behaviors to predict user post-click conversion in an end-to-end approach. Specifically, we define a general aggregation operator to combine predicted output of different tasks to a unified score, then we use the computed scores to model the label relations. Inspired by recent advances on differentiable sorting, we seamlessly integrated a label sorting component into the MTL structure so they can be simultaneously optimized in training. We call the model \textbf{\PM{}} which reveals the two main techniques: \textbf{M}ulti-\textbf{T}ask \textbf{L}earning and \textbf{D}ifferentiable \textbf{S}orting. Extensive experiments on public and industrial datasets show the superiority of our proposed framework against competitive baselines.

\section{Related Work}
Post-click conversion prediction is widely explored in many online applications. Many efforts are devoted on the sample selection bias and sparsity problem.
. \citet{ma2018entire} proposed an entire space multi-task model for predicting CVR, which remedies the data sparsity
problem. Further, \citet{wen2020entire} leveraged supervisory signals from users’ post-click behaviors other than conversions to further alleviate the data sparsity problem. \citet{bao2020gmcm} utilized graph convolutional networks (GCN) to enhance the conventional CVR modeling. However, these methods all belong to pointwise learning.

Another branch of related work is differentiable sorting. Prior works \cite{mena2018learning,linderman2018reparameterizing} have proposed relaxations of permutation matrices to the Birkhoff polytope, which is defined as the convex hull of the set of doubly-stochastic matrices. Recent works \cite{grover2018stochastic,prillo2020softsort} mapped permutation matrices to the set of unimodal row-stochastic matrices and then perform \texttt{softmax} operation to model the sorted index of permutation matrices. 


\section{Framework}

\begin{figure}[t]
\centering 
\includegraphics[width=5cm]{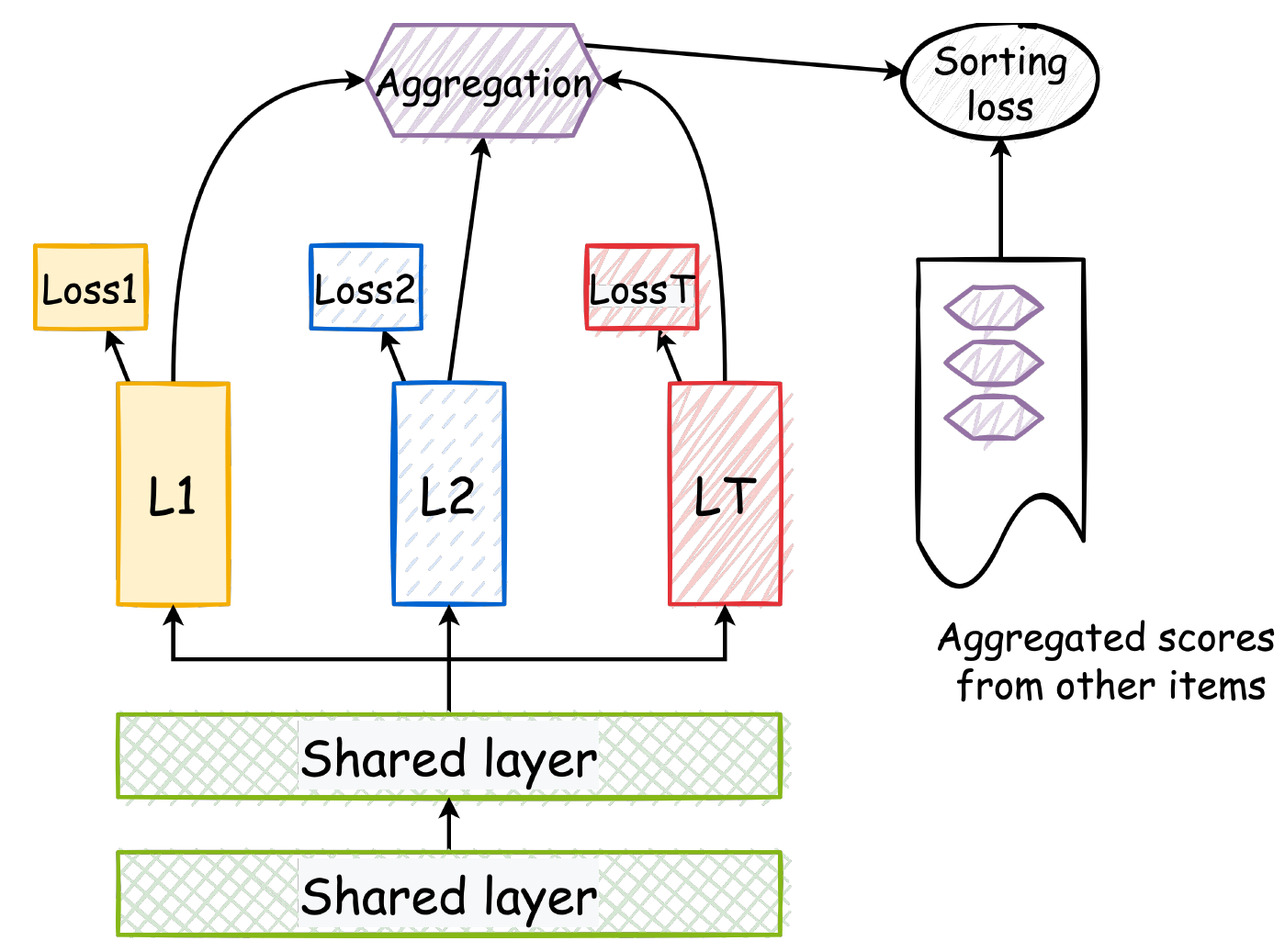}
\caption{The structure overview of proposed framework. }
\label{model}
\end{figure}

Assuming we have a dataset $\{X,L\}$, where $X$ is feature set and $L$ is the ordered label sequence of length $T$. Concretely, $L = (l_1, l_2,..., l_T)$; $L_t \in \{0,1\}, \forall t\in[1,T]$. In other words, each label is a binary variable. For post-click behaviors, we denote those labels with index $t>L_{click}$
If our target is click, the problem degenerates to click-through rate prediction.
For simplicity and clarity, we set $T=2$ in the remaining part of this chapter, and we denote $L_{click}=L_1$ and $L_{post_click}=L_2$. Note that our framework applies to arbitrary length of $T$.

Regarding the sorting part, we follow the definition of \citet{grover2018stochastic}. An $n$-dimensional permutation $\mathbf{z} = [z_1, z_2, \ldots, z_n]^T$ is a list of unique indices $\{1,2,\ldots,n\}$. 
Every permutation $\mathbf{z}$ is associated with a permutation matrix $P_\mathbf{z}\in {\{0,1\}^{n \times n}}$ with entries given as:
\begin{align*}
P_\mathbf{z}[i,j] = 
\begin{cases}
1 \text{ if } j = z_i\\
0 \text{ otherwise}.
\end{cases}
\end{align*} 

This is simply the one-hot representation of $\pi$.
Note that with these definitions, the mapping $\texttt{sort} : \mathbb R^n \rightarrow \mathbb R^n$ that sorts $s$ in decreasing order is $\texttt{sort}(s) = P_{\texttt{argsort}(s)}s$. The differentiable sorting in this paper mainly indicates making $P_{\texttt{argsort}(s)}$ differentiable.

\subsection{Architecture overview.}
We choose a multi-task structure as the backbone of our framework \PM{}. As shown form Fig. \ref{model}, we adopt hard parameter sharing to common bottom layers, and a task-specific neural network for each task. Hard parameter sharing is the most commonly used approach to multi-task learning in neural networks, which has been empirically proved to reduce the risk of fitting. On top of the backbone, we have an \textbf{aggregation module} and a \textbf{sorting module}. The aggregation module aggregates outputs and labels of each task into a unified score and label, and then feeds them into the sorting module. Next, the sorting module will rank the orders of labels and predictions, which yields a sorting loss. Finally, the total loss function becomes 
 \begin{equation}\label{eq:loss}
L = \sum_t^T Loss_t + Loss_\texttt{sort},
\end{equation}
where $Loss_t$ is task-specific loss.

\subsection{Aggregations.} Here we introduce and discuss two sources utilized in the aggregation modular: predictions and labels of multi-tasks. 1) \emph{Prediction aggregation.}
Without loss of generality, we define an \emph{aggregation} operator $g(\cdot)$ to combine predictions of multiple tasks. So for a predicted probability sequence $\hat{L}$ we can compute a score $s = g(\hat{L})$. We implemented the following three candidates:
\begin{itemize}
  \item \textbf{Mul}-operator. This idea originates from conditional probability decomposition, which assumes post-click behaviors are conditioned on clicks,. a widely adopted idea. One drawback is the probability decomposition is difficult to get when $T$ is very large. $g = \prod_{t}^{T}\hat{L}_t$.
  \item \textbf{Max}-operator. $s$ is assigned as the maximum value of $\hat{L}$, i.e., $g = max(\hat{L})$.
  \item \textbf{Sum}-operator. This operator assumes $g = \sum_{t}^{T}w_t\hat{L}_t$, where $w_t$ is a task-specific weight. In experiments we employ two variants: \textbf{Add} and \textbf{Linear}. The former sets all $w_t$ to $1$, and the latter sets $w_t$ as learnable parameters.  

\end{itemize}

\emph{Discussion} The above three operators come from different assumptions with respect to data distributions and user behaviors. We provide an explainable example in Table \ref{tb:example} to demonstrate their difference and applicability.
\textbf{Mul}-operator has a relaxed mapping to probability decomposition: 
 \begin{equation*}\label{eq:loss}
P(\textit{Click}, \textit{Post-click}) = P(\textit{Click}) * P(\textit{Post-click} | \textit{Click}),
\end{equation*}
which is widely adopted by previous work. However, multiplication may not well discriminate some cases, e.g., \texttt{Sample 1-3} in Table \ref{tb:example} achieve the same aggregated score $0.09$, while the raw $P(\textit{Click})$ and $P(\textit{Post-click})$ are completely different. Meanwhile, \textbf{Mul}-operator implicitly suffers from a high variance problem. $P(\textit{Click})$ and $P(\textit{Post-click})$ need to be calibrated to ensure an accurate multiplication. \textbf{Max}-operator, which equals to the one-dimension global max pooling layer, extracts the most prominent output from multiple tasks. In this regard, we can assume user actions on one item are driven by the highest probability. In contrast, \textbf{Sum}-operator provides a smooth approach to linearly consider the relations between different labels. It has the most moderate assumption thus it can tell the difference between \textit{Sample 1-2} with \textit{w 3:2} (weights), which \textbf{Mul}-operator and \textbf{Max}-operator fail to distinguish.

2) \emph{Label aggregation.} We adopt a simple but reasonable approach to merge labels. For a single sample, we sum all the binary labels of multiple tasks to a score $s$. Then for samples under the same impression, we use computed $s$ to form a permutation matrix. The permutation matrix is served as the label for the sorting loss. Note that exploring orders of user actions across samples is meaningful.

\begin{table}[t]
    \scriptsize
	\centering
	\caption{Explainable example of three aggregation operators.}
	\label{tb:example}
	\resizebox{\linewidth}{!}{
	\begin{tabular}{c|cc|c|ccc}
		\toprule
		\vtop{\hbox{\strut Sample}\hbox{\strut Index}}      & P(Click) & P(Post-click) & Mul  & \vtop{\hbox{\strut Sum}\hbox{\strut (w 1:1)}} &  \vtop{\hbox{\strut Sum}\hbox{\strut (w 3:2)}} & Max        \\ \midrule
		
		1     & 0.9 & 0.1  & 0.09 &  1.0  & 2.9 &  0.9     \\ \midrule
	   	2     & 0.1 & 0.9  & 0.09 &  1.0  &  2.1 &  0.9      \\ \midrule
		3      & 0.3  & 0.3  & 0.09 & 0.6  &  1.5   &  0.3    \\ \midrule
		4  & 0.5  & 0.5  & 0.25   & 1.0  & 2.5  & 0.5    \\
	   \bottomrule
	\end{tabular}
	}
\end{table}

\subsection{Learning orders of user actions.}
After receiving prediction sequence $s$ and permutation matrix $z$ from the aggregation modular, we turn to computing permutation probability via
mapping prediction matrices to a set of unimodal row-stochastic matrices. Here we apply softsort \cite{prillo2020softsort}, a simple but efficient implementation of differentiable sorting. Concretely,

\begin{equation}\label{eq:simplesort}
\hat{P}_z = \texttt{softmax}\left(\frac{-d\left(\texttt{sort}(s) \mathds 1^T, \mathds 1 s^T\right)}{\tau}\right)
\end{equation}
where the softmax operator is applied row-wise, $d$ is an element-wise Manhattan Distance function, and $\tau$ is a temperature parameter that controls the degree of the approximation. This sorting operation relaxes permutation matrices to a set of unimodal row-stochastic matrices (the former is row-stochastic and column-stochastic). The definition of \textit{Unimodal Row Stochastic Matrices} \citet{grover2018stochastic} summarizes the properties of our relaxed operator:

\begin{definition} (Unimodal Row Stochastic Matrices).
An $n \times n$ matrix is Unimodal Row Stochastic (URS) if it satisfies the following conditions:
\begin{enumerate}
\item \textbf{Non-negativity:} $U[i, j] \ge 0\quad \forall i,j \in \{1,2,\dots,n\}$.
\item \textbf{Row Affinity:} $\sum_{j = 1}^n U[i, j] = 1\quad \forall i \in \{1,2,\dots,n\}$.
\item \textbf{Argmax Permutation:} Let $u$ denote a vector of size $n$ such that $u_i = \arg\max_j U[i, j]\quad \forall i \in \{1,2,\dots,n\}$. Then, $u \in \mathcal S_n$, i.e., it is a valid permutation.
\end{enumerate}
\end{definition}
All row stochastic matrices satisfy the first two conditions. The third condition is useful for gradient based optimization involving sorting-based losses.
The whole proof and discussion of can be found in \citet{prillo2020softsort} and \citet{grover2018stochastic}. In simple words: \textit{the $r$-th row of the operator is the \texttt{softmax} of the negative distances to the $r$-th largest element}.

We then minimized the cross-entropy loss between the predicted matrix $\hat{P}_z$ and the ground-truth permutation matrix $P_z$, concretely,
\begin{equation}
Loss_\texttt{sort}  = - \displaystyle\sum_{i=1}^n w_i \sum_{j=1}^n    P_{z_{ij}} \log(\hat{P}_{z_{ij}}) +  (1 - P_{z_{ij}}) \log(1-\hat{P}_{z_{ij}}),
\label{eq:sort_loss}
\end{equation}
where $w_i$ is penalty weight at position index $i$. Similar to the setting of NDCG, we apply $w_i=\frac{1}{\log_2(i+1)}$ and find it useful in practice.

\section{Experiments}\label{sec:exp}
\textbf{Datasets} We selected two datasets for evaluating recommendation performance, one public dataset and one industrial dataset. For the public dataset\footnote{http://yongfeng.me/dataset/}, we choose the Alibaba E-commerce user behavior dataset. For the industrial dataset, we crawled $7$ days' data from one placement in an e-commerce website, with click, addtoCart, and purchase labels. The statistics are detailed in Table \ref{tb:data_statistics}.

\begin{table}[h]
   \footnotesize
	\centering
	\caption{Statistics of datasets (size $\times 10^6$).}
	\label{tb:data_statistics}
	\vspace{-2ex}
	\resizebox{\linewidth}{!}{
	\begin{tabular}{c|cc|c|ccc}
		\toprule
		Dataset      & $\#$User & $\#$Item & $\#$Sample  &  $\#$Click &  $\#$Purchase & $\#$Add2Cart        \\ \midrule
		
		Taobao    & 49 & 200  & 2500 &  670  & 3 &  -    \\ \midrule
		\DT{}  & 0.31  & 1.2  & 16.8   & 0.1  & 0.008  & 0.01    \\
	   \bottomrule
	\end{tabular}
	}
\end{table}

 \begin{table*}[ht]
  \footnotesize
        \caption{Comparisons of different models on two datasets.}
        \label{tab:offexp}
        \vspace{-1ex}\begin{tabular}{l|c|c|c|c|c|c|c|c}
            \toprule
            \multirow{2}{*}{Models} &  \multicolumn{4}{c|}{Taobao}              &                \multicolumn{4}{c}{\DT{}}          \\
            \cline{2-9}
                           & AUC & NDCG@2 & NDCG@6 & NDCG@12 & AUC & NDCG@2 & NDCG@6 & NDCG@12 \\ 
            \midrule
           DNN-Pointwise          & 0.5210 & 0.0918 & 0.2399 & 0.4110     & 0.5837 & 0.1027 & 0.1542 & 0.3724 \\
           DNN-Pairwise          & 0.5122 & 0.2261 & 0.3989 & 0.5085               & 0.7332 & 0.5401 & 0.6857 & 0.7042 \\
           DNN-DiffSort & 0.5622 & 0.2682 & 0.4288 & 0.5347                 & 0.7289 & 0.5455 & 0.6861 & 0.7026 \\
           ESMM         & 0.5268 & 0.2286 & 0.3986 & 0.5101                 & 0.4329 & 0.3323 & 0.5322 & 0.5793 \\
           ESMM-Pairwise    & 0.5383 & 0.2337 & 0.4048 & 0.5147                 & 0.7255 & 0.5355 &	0.6831 & 0.7018 \\
           \midrule
           MTL-Linear-ListNet    & 0.5213 & 0.2580 & 0.4358 & 0.5319                 & 0.7116 & 0.5277 & 0.6771 & 0.6963 \\
           MTL-Linear-NeuralNDCG    & 0.5153 & 0.2430 & 0.4171 & 0.5196                 & 0.7128 & 0.5331 & 0.6819 & 0.7006 \\
           \midrule
           \PM -Max & 0.5720 & 0.2913 & 0.4677 & 0.5540     & 0.7349 & 0.5477 & 0.6920 & 0.7096 \\ 
           \PM -Mul & 0.5724 & 0.2861 &	0.4613 & 0.5496     & 0.7327 & 0.5435 & 0.6878 & 0.7058 \\ 
           \PM -Add & 0.5615 & 0.2774 & 0.4468 & 0.5419     & 0.7335 & 0.5499 & 0.6909 & 0.7088 \\ 
            
           \PM -Linear & \textbf{0.5844} & \textbf{0.2995} & \textbf{0.4693} & \textbf{0.5563}                   & \textbf{0.7418} & \textbf{0.5558} & \textbf{0.6962} & \textbf{0.7129} \\ 
            \bottomrule

        \end{tabular}
\end{table*}

\noindent\textbf{Baselines} We conduct experiments with several competitive methods on CVR modeling. 1) \textbf{DNN-pointwise}~\cite{naumov2019deep}. This is a representative single-task model with point-wise binary cross-entropy loss where we only utilize purchase as labels. 2) \textbf{DNN-pairwise.} In this model, we construct a combined label (0, 1, 2) by adding click label to purchase label and implement paiwise logistic loss. 3) \textbf{DNN-DiffSort}. This model implements the proposed listwise sorting loss with the combined label as used in \textbf{DNN-pairwise}. 4) \textbf{ESMM}~\cite{ma2018entire}. This is a multi-task DNN model simultaneously predicting the CTR, CVR and their multiplication CTCVR with shared user/item embeddings. 5) \textbf{ESMM-Pairwise}. This model replaces the pointwise binary cross-entropy loss used in ESMM to ranknet loss~\cite{burges2005learning}. 6) \textbf{MTL-Linear-ListNet}. This model follows the proposed user behavior order modeling multi-task structure with its loss function replaced by listNet ranking loss ~\cite{cao2007learning} and linear aggregator. 7) \textbf{MTL-Linear-NeuralNDCG}. This model also follows the proposed multi-task structure using NeuralNDCG loss ~\cite{pobrotyn2021neuralndcg} and linear aggregator. Note that in all MTL models, we apply ranknet loss~\cite{burges2005learning} to domain-specific task losses, which we find competitive.

\noindent\textbf{Relation to listwise learning}. Differentiable sorting can also be viewed as list-wise learning. So we include two list-wise loss into \PM{} for comparison. 1) ListNet \cite{cao2007learning} utilizes SoftMax method to project both predicted scores and labels into probability space and then minimizes the cross-entropy loss between them. In other words, ListNet generates soft labels (i.e., \emph{softmax} output of list labels) for model learning, which is an implicit mehtod for listwise learning.
2) NeuralNDCG \cite{pobrotyn2021neuralndcg} is a differentiable approximated Normalised Discounted Cumulative Gain (NDCG) loss based on another differentiable sorting function: Neuralsort~\cite{grover2018stochastic}. The sorting approximation relies on an identity that expresses the sum of the top $k$ elements of a vector $s \in \mathbb R^n$ as a symmetric function of $s$ that only involves \texttt{max} and \texttt{min} operations~\citep[Lemma 1]{ogryczak2003minimizing}. Comparing with the above two counterparts, our proposed \PM{} model is a more direct way to optimize the ranking ability across the whole list, and easier to implement and optimize. Moreover, \PM{} is extended to multi-task learning with a general aggregation operator. 

 \begin{table}[ht]
        \scriptsize
        \caption{Model performance with/without AddtoCart label.}
        \label{tab:offexp2}
        \begin{tabular}{l|c|c|c}
            \toprule
            \multirow{2}{*}{Models} &  \multicolumn{3}{c}{\DT{}}   \\
            \cline{2-4}
                           & NDCG@2 & NDCG@6 & NDCG@12\\ 
           \midrule
           \PM -Max  & 0.5477 & 0.6920 & 0.7096 \\
           \PM -Max-Cart  & \textbf{0.5501} & \textbf{0.6930} & \textbf{0.7103} \\
           \midrule
           \PM -Mul  & 0.5435 & 0.6878 & 0.7058 \\
           \PM -Mul-Cart  & \textbf{0.5492} & \textbf{0.6926} & \textbf{0.7099} \\
           \midrule
           \PM -Add  & 0.5499 & 0.6909 & 0.7088 \\
           \PM -Add-Cart  & \textbf{0.5528} & \textbf{0.6945} & \textbf{0.7118} \\
           \midrule
           \PM -Linear  & {0.5558} & {0.6962} & {0.7129} \\
           \PM -Linear-Cart  & \textbf{0.5601} & \textbf{0.6992} & \textbf{0.7163} \\
            \bottomrule
        \end{tabular}
\end{table}

\noindent\textbf{Experimental results of purchase prediction}. A common practice of post-click conversion prediction is purchase prediction. Here we assume label orders follows \emph{Click -> Purchase} and conduct experiments on the aforementioned two datasets. We report model performance in terms of AUC, NDCG@2. NDCG@6, and NDCG@12 of purchase prediction. Results are collected in Table \ref{tab:offexp}, from which we can reveal following findings: 1) \PM{} variants outperform other competitors consistently on all metrics across both datasets, showing the superiority of our proposed model. 2) Models with listwise loss are better than those with pairwise loss or pointwise loss, e.g., DNN-Diffsort vs DNN-Purchase and DNN-Pairwise, ESMM-Pairwise vs ESMM. These results meet expectation as it is widely discussed in literature~\cite{liu2011learning}. 3) Comparing among three listwise losses (i.e., ListNet, NeuralNDCG, and \PM-Linear), we observe a considerable improvement of the differentiable sorting function against other two losses. This finding implies that directly estimating the permutation matrix is a better way. 4) Regarding the four different aggregators in \PM{}, the linear aggregator performs the best on two datasets consistently. This confirms the previous demonstration that \textbf{Mul}-operator, \textbf{Max}-operator and vanilla \textbf{Sum}-operator could fail to distinguish certain user behavior sequences. The \textbf{Linear}-operator has the best scoring ability.

\noindent\textbf{Experimental results of multiple post-click labels}.
To verify \PM's scalability with more than one post-click labels, we introduce another label \textit{AddtoCart} and assume a \emph{Click -> AddtoCart -> Purchase} ordering for user behavior. Results in Table \ref                           {tab:offexp2} show that incorporating \textit{AddtoCart} labels helps improve the performance of \PM{} with all aggregation operators in purchase prediction. Note that we don't have to design additional loss functions or aggregation operators for additional post-click labels. In this regard, we argue \PM{} is scalable and general.

\section{Conclusion}
In this paper we propose a novel multi-task framework \PM{} that leverages orders of user
behaviors to predict user post-click conversion in an end-to-end approach. 
Concretely, we define an aggregation operator to combine predicted output of different tasks to a unified score, then we use the computed scores to model the label relations via differentiable sorting.
Extensive experiments on public and industrial datasets show the superiority of our proposed model against competitive baselines. In the future, we plan to explore \PM{} with graph-based user behaviors.

\bibliographystyle{ACM-Reference-Format}
\bibliography{mtl_sorting}


\begin{thebibliography}{14}


\ifx \showCODEN    \undefined \def \showCODEN     #1{\unskip}     \fi
\ifx \showDOI      \undefined \def \showDOI       #1{#1}\fi
\ifx \showISBNx    \undefined \def \showISBNx     #1{\unskip}     \fi
\ifx \showISBNxiii \undefined \def \showISBNxiii  #1{\unskip}     \fi
\ifx \showISSN     \undefined \def \showISSN      #1{\unskip}     \fi
\ifx \showLCCN     \undefined \def \showLCCN      #1{\unskip}     \fi
\ifx \shownote     \undefined \def \shownote      #1{#1}          \fi
\ifx \showarticletitle \undefined \def \showarticletitle #1{#1}   \fi
\ifx \showURL      \undefined \def \showURL       {\relax}        \fi
\providecommand\bibfield[2]{#2}
\providecommand\bibinfo[2]{#2}
\providecommand\natexlab[1]{#1}
\providecommand\showeprint[2][]{arXiv:#2}

\bibitem[Bao et~al\mbox{.}(2020)]%
        {bao2020gmcm}
\bibfield{author}{\bibinfo{person}{Wentian Bao}, \bibinfo{person}{Hong Wen},
  \bibinfo{person}{Sha Li}, \bibinfo{person}{Xiao-Yang Liu},
  \bibinfo{person}{Quan Lin}, {and} \bibinfo{person}{Keping Yang}.}
  \bibinfo{year}{2020}\natexlab{}.
\newblock \showarticletitle{Gmcm: Graph-based micro-behavior conversion model
  for post-click conversion rate estimation}. In
  \bibinfo{booktitle}{\emph{Proceedings of the 43rd International ACM SIGIR
  Conference on Research and Development in Information Retrieval}}.
  \bibinfo{pages}{2201--2210}.
\newblock


\bibitem[Burges et~al\mbox{.}(2005)]%
        {burges2005learning}
\bibfield{author}{\bibinfo{person}{Chris Burges}, \bibinfo{person}{Tal Shaked},
  \bibinfo{person}{Erin Renshaw}, \bibinfo{person}{Ari Lazier},
  \bibinfo{person}{Matt Deeds}, \bibinfo{person}{Nicole Hamilton}, {and}
  \bibinfo{person}{Greg Hullender}.} \bibinfo{year}{2005}\natexlab{}.
\newblock \showarticletitle{Learning to rank using gradient descent}. In
  \bibinfo{booktitle}{\emph{Proceedings of the 22nd international conference on
  Machine learning}}. \bibinfo{pages}{89--96}.
\newblock


\bibitem[Cao et~al\mbox{.}(2007)]%
        {cao2007learning}
\bibfield{author}{\bibinfo{person}{Zhe Cao}, \bibinfo{person}{Tao Qin},
  \bibinfo{person}{Tie-Yan Liu}, \bibinfo{person}{Ming-Feng Tsai}, {and}
  \bibinfo{person}{Hang Li}.} \bibinfo{year}{2007}\natexlab{}.
\newblock \showarticletitle{Learning to rank: from pairwise approach to
  listwise approach}. In \bibinfo{booktitle}{\emph{Proceedings of the 24th
  international conference on Machine learning}}. \bibinfo{pages}{129--136}.
\newblock


\bibitem[Grover et~al\mbox{.}(2018)]%
        {grover2018stochastic}
\bibfield{author}{\bibinfo{person}{Aditya Grover}, \bibinfo{person}{Eric Wang},
  \bibinfo{person}{Aaron Zweig}, {and} \bibinfo{person}{Stefano Ermon}.}
  \bibinfo{year}{2018}\natexlab{}.
\newblock \showarticletitle{Stochastic Optimization of Sorting Networks via
  Continuous Relaxations}. In \bibinfo{booktitle}{\emph{International
  Conference on Learning Representations}}.
\newblock


\bibitem[Linderman et~al\mbox{.}(2018)]%
        {linderman2018reparameterizing}
\bibfield{author}{\bibinfo{person}{Scott Linderman}, \bibinfo{person}{Gonzalo
  Mena}, \bibinfo{person}{Hal Cooper}, \bibinfo{person}{Liam Paninski}, {and}
  \bibinfo{person}{John Cunningham}.} \bibinfo{year}{2018}\natexlab{}.
\newblock \showarticletitle{Reparameterizing the birkhoff polytope for
  variational permutation inference}. In
  \bibinfo{booktitle}{\emph{International Conference on Artificial Intelligence
  and Statistics}}. PMLR, \bibinfo{pages}{1618--1627}.
\newblock


\bibitem[Liu(2011)]%
        {liu2011learning}
\bibfield{author}{\bibinfo{person}{Tie-Yan Liu}.}
  \bibinfo{year}{2011}\natexlab{}.
\newblock \showarticletitle{Learning to rank for information retrieval}.
\newblock  (\bibinfo{year}{2011}).
\newblock


\bibitem[Ma et~al\mbox{.}(2018)]%
        {ma2018entire}
\bibfield{author}{\bibinfo{person}{Xiao Ma}, \bibinfo{person}{Liqin Zhao},
  \bibinfo{person}{Guan Huang}, \bibinfo{person}{Zhi Wang},
  \bibinfo{person}{Zelin Hu}, \bibinfo{person}{Xiaoqiang Zhu}, {and}
  \bibinfo{person}{Kun Gai}.} \bibinfo{year}{2018}\natexlab{}.
\newblock \showarticletitle{Entire space multi-task model: An effective
  approach for estimating post-click conversion rate}. In
  \bibinfo{booktitle}{\emph{The 41st International ACM SIGIR Conference on
  Research \& Development in Information Retrieval}}.
  \bibinfo{pages}{1137--1140}.
\newblock


\bibitem[Mena et~al\mbox{.}(2018)]%
        {mena2018learning}
\bibfield{author}{\bibinfo{person}{Gonzalo Mena}, \bibinfo{person}{David
  Belanger}, \bibinfo{person}{Scott Linderman}, {and} \bibinfo{person}{Jasper
  Snoek}.} \bibinfo{year}{2018}\natexlab{}.
\newblock \showarticletitle{Learning latent permutations with gumbel-sinkhorn
  networks}.
\newblock \bibinfo{journal}{\emph{arXiv preprint arXiv:1802.08665}}
  (\bibinfo{year}{2018}).
\newblock


\bibitem[Naumov et~al\mbox{.}(2019)]%
        {naumov2019deep}
\bibfield{author}{\bibinfo{person}{Maxim Naumov}, \bibinfo{person}{Dheevatsa
  Mudigere}, \bibinfo{person}{Hao-Jun~Michael Shi}, \bibinfo{person}{Jianyu
  Huang}, \bibinfo{person}{Narayanan Sundaraman}, \bibinfo{person}{Jongsoo
  Park}, \bibinfo{person}{Xiaodong Wang}, \bibinfo{person}{Udit Gupta},
  \bibinfo{person}{Carole-Jean Wu}, \bibinfo{person}{Alisson~G Azzolini},
  {et~al\mbox{.}}} \bibinfo{year}{2019}\natexlab{}.
\newblock \showarticletitle{Deep learning recommendation model for
  personalization and recommendation systems}.
\newblock \bibinfo{journal}{\emph{arXiv preprint arXiv:1906.00091}}
  (\bibinfo{year}{2019}).
\newblock


\bibitem[Ogryczak and Tamir(2003)]%
        {ogryczak2003minimizing}
\bibfield{author}{\bibinfo{person}{Wlodzimierz Ogryczak} {and}
  \bibinfo{person}{Arie Tamir}.} \bibinfo{year}{2003}\natexlab{}.
\newblock \showarticletitle{Minimizing the sum of the k largest functions in
  linear time}.
\newblock \bibinfo{journal}{\emph{Inform. Process. Lett.}}
  \bibinfo{volume}{85}, \bibinfo{number}{3} (\bibinfo{year}{2003}),
  \bibinfo{pages}{117--122}.
\newblock


\bibitem[Pobrotyn and Bia{\l}obrzeski(2021)]%
        {pobrotyn2021neuralndcg}
\bibfield{author}{\bibinfo{person}{Przemys{\l}aw Pobrotyn} {and}
  \bibinfo{person}{Rados{\l}aw Bia{\l}obrzeski}.}
  \bibinfo{year}{2021}\natexlab{}.
\newblock \showarticletitle{NeuralNDCG: Direct Optimisation of a Ranking Metric
  via Differentiable Relaxation of Sorting}.
\newblock \bibinfo{journal}{\emph{arXiv preprint arXiv:2102.07831}}
  (\bibinfo{year}{2021}).
\newblock


\bibitem[Prillo and Eisenschlos(2020)]%
        {prillo2020softsort}
\bibfield{author}{\bibinfo{person}{Sebastian Prillo} {and}
  \bibinfo{person}{Julian Eisenschlos}.} \bibinfo{year}{2020}\natexlab{}.
\newblock \showarticletitle{SoftSort: A continuous relaxation for the argsort
  operator}. In \bibinfo{booktitle}{\emph{International Conference on Machine
  Learning}}. PMLR, \bibinfo{pages}{7793--7802}.
\newblock


\bibitem[Wen et~al\mbox{.}(2020)]%
        {wen2020entire}
\bibfield{author}{\bibinfo{person}{Hong Wen}, \bibinfo{person}{Jing Zhang},
  \bibinfo{person}{Yuan Wang}, \bibinfo{person}{Fuyu Lv},
  \bibinfo{person}{Wentian Bao}, \bibinfo{person}{Quan Lin}, {and}
  \bibinfo{person}{Keping Yang}.} \bibinfo{year}{2020}\natexlab{}.
\newblock \showarticletitle{Entire space multi-task modeling via post-click
  behavior decomposition for conversion rate prediction}. In
  \bibinfo{booktitle}{\emph{Proceedings of the 43rd International ACM SIGIR
  conference on research and development in Information Retrieval}}.
  \bibinfo{pages}{2377--2386}.
\newblock


\bibitem[Zadrozny(2004)]%
        {zadrozny2004learning}
\bibfield{author}{\bibinfo{person}{Bianca Zadrozny}.}
  \bibinfo{year}{2004}\natexlab{}.
\newblock \showarticletitle{Learning and evaluating classifiers under sample
  selection bias}. In \bibinfo{booktitle}{\emph{Proceedings of the twenty-first
  international conference on Machine learning}}. \bibinfo{pages}{114}.
\newblock


\end{thebibliography}
\end{document}